\documentclass[prl,aps,twocolumn,superscriptaddress,showpacs]{revtex4}

\usepackage{pxfonts}
\usepackage{graphicx}
\begin{document}

\title{Spin transport of indirect excitons in GaAs coupled quantum wells}

\author{J.R. Leonard}
\author{Sen Yang}
\author{L.V. Butov}
\affiliation{Department of Physics, University of California at San
Diego, La Jolla, CA 92093-0319}

\author{A.C. Gossard}
\affiliation{Materials Department, University of California at Santa
Barbara, Santa Barbara, California 93106-5050}

\begin{abstract}
Spin transport of indirect excitons in GaAs/AlGaAs coupled quantum
wells was observed by measuring the spatially resolved circular
polarization of the exciton emission. The exciton spin transport
originates from the long spin relaxation time and long lifetime of
the indirect excitons.
\end{abstract}


\date{\today}

\maketitle

Spin physics in semiconductors includes a number of interesting
phenomena in electron transport, such as current-induced
spin orientation (the spin Hall effect)
\cite{Dyakonov1971,Hirsch1999,Sih2005}, spin-induced contribution to
the current \cite{Dyakonov1971b}, spin injection \cite{Aronov1976},
and spin diffusion and drag
\cite{Kikkawa1999,Amico2001,Weber2005,Carter2006}. Besides the
fundamental spin physics, there is also considerable interest in
developing semiconductor electronic devices based on spin
transport, which may offer advantages in dissipation, size and speed
over charge-based devices, see \cite{Wolf2000,Awschalom2007} and
references therein.

Optical methods have been used as a tool for precise probe and
control of electron spin via photon polarization and, in
particular, for studying electron spin transport in
semiconductors \cite{Wolf2000,Stievater2001,Awschalom2007}. Excitons
play a major role in the optical properties of quantum wells (QW)
near the fundamental absorption edge. The spin dynamics of excitons
in GaAs/AlGaAs single QW was extensively studied in the past, see
\cite{Maialle1993,Vinattieri1994} and references therein. It was
found that the spin relaxation time of the excitons in single QW is
typically limited by the electron-hole exchange interaction and is
short, on the order of a few tens of ps. Because of the short spin
relaxation time, no spin transport of excitons was observed in
single GaAs QWs.

Here, we report on the spin transport of indirect excitons in GaAs
coupled quantum wells (CQW). The spin relaxation time of the
indirect excitons is orders of magnitude longer than for regular
excitons in single QW. In combination with the long lifetime of the
indirect excitons, this makes possible the spin transport of the
indirect excitons over substantial distances (up to several microns
in the present work).

The spin dynamics of an exciton is related to the spin dynamics of
its constituents - an electron and a hole
\cite{Andreani1990,Maialle1993,Vinattieri1994}. The $\sigma^+$
($\sigma^-$) polarized light propagating along the $z$-axis creates
a heavy hole exciton with the electron spin state $s_z=-1/2$
($s_z=+1/2$) and hole spin state $m_h=+3/2$ ($m_h=-3/2$) in GaAs QW
structures. Then the exciton spin $S_z=s_z+m_h$ can be changed by
flipping the spin of either the electron or the hole or by
simultaneous flipping the spin of both electron and hole. In
high-quality samples with a small concentration of magnetic
impurities, the electron spin relaxation is typically determined by
the Dyakonov-Perel \cite{Dyakonov1971a} mechanism. The hole spin
relaxation in GaAs structures is typically determined by the hole
momentum relaxation because a valence state is a mixture of the
$z$-components of the 3/2 spin \cite{Uenoyama1990}. The exciton spin
relaxation by simultaneous flipping the spin of both electron and
hole is driven by the electron-hole exchange interaction
\cite{Andreani1990,Maialle1993,Vinattieri1994}.

Heavy hole excitons with $S_z=+1$ (-1) emit $\sigma^+$ ($\sigma^-$)
polarized light while excitons with $S_z=\pm 2$ are optically
inactive. The dynamics of the polarization of the exciton emission
$P=(I_+ - I_-)/(I_+ + I_-)$ is determined by the recombination and
spin relaxation processes. For an optically active heavy hole
exciton with spin $S_z=\pm 1$, spin flip of either electron or hole
would transform the exciton to an optically inactive state with spin
$S_z=\pm 2$, Fig. 1a. Therefore, these processes do not cause the
decay of the emission polarization. The polarization decays when
both the electron and hole flip their spins. This occurs in the
two-step process due to the separate electron and hole spin flips
and the single-step process due to the simultaneous flipping the
spin of both electron and hole. The rate equations describing these
processes \cite{Maialle1993,Vinattieri1994} yield the polarization
of the exciton emission $P=\tau_p/(\tau_p+\tau_r)$, where
$\tau_p^{-1}=2(\tau_e+\tau_h)^{-1}+\tau_{ex}^{-1}$ is the relaxation
time of the emission polarization, $\tau_{ex}$ is time for exciton
flipping between $S_z= \pm 1$, $\tau_e$ and $\tau_h$ are electron
and hole spin flip times, and $\tau_r$ is the exciton recombination
time (for the typical case when the energy splitting between
$S_z=\pm 1$ and $S_z=\pm 2$ states due to the electron-hole exchange
interaction is smaller than $k_BT$).

\begin{figure}
\begin{center}
\includegraphics[width=7cm]{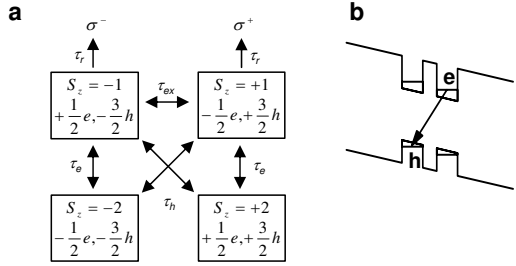}
\caption{\label{fig:fig1}(a) Exciton spin diagram, see text. (b)
Energy diagram of the CQW structure; e, electron; h, hole.}
\end{center}
\end{figure}

In GaAs single QW, $\tau_h$ and $\tau_{ex}$ are typically short, in
the range of several tens of ps, while $\tau_e$ can reach tens and
hundreds of ns in high-quality samples. It is the long electron spin
relaxation time, which makes possible the spin transport of
electrons over large distances, see
\cite{Kikkawa1999,Weber2005,Carter2006,Wolf2000,Awschalom2007}. For
$\tau_e \gg \tau_h, \tau_{ex}$, which is typical of GaAs single QW,
$\tau_p \approx \tau_{ex}$ and, therefore, the small $\tau_{ex}$
results in a fast depolarization of the exciton emission within tens
of ps \cite{Maialle1993,Vinattieri1994}. However, $\tau_{ex}$ is
determined by the strength of the exchange Coulomb interaction
between the electron and hole and is inversely proportional to the
electron-hole overlap. This gives an opportunity to control the
depolarization rate by changing the electron-hole overlap, e.g. in
QW structures with different QW widths or with an applied electric
field \cite{Maialle1993,Vinattieri1994}.

The electron-hole overlap is drastically reduced in CQW structures.
An indirect exciton in CQW is composed of an electron and a hole
confined in spatially separated quantum wells, Fig. 1b. The spatial
separation results in a small electron-hole overlap. One of the
results of such small electron-hole overlap is a long exciton
recombination time $\tau_r$, which is typically in the range between
tens of ns to tens of $\mu s$
\cite{Alexandrou1990,Zrenner1992,Butov1999}, orders of magnitude
longer than $\tau_r$ in single QW, which is typically in the range
of tens and hundreds of ps \cite{Maialle1993,Vinattieri1994}. The
long lifetimes of the indirect excitons make possible their
transport over large distances, which can reach tens and hundreds of
microns
\cite{Hagn1995,Larionov2000,Butov2002,Voros2005,Ivanov2006,Gartner2006}.
The small electron-hole overlap for the indirect excitons should
also result in a large $\tau_{ex}$ and, in turn, $\tau_p$, thus
making possible the exciton spin transport over large distances.

We probed exciton spin transport in a GaAs/AlGaAs CQW structure.
Two 8 nm GaAs QW were separated by a 4 nm Al$_{0.33}$Ga$_{0.67}$As
barrier and surrounded by 200 nm Al$_{0.33}$Ga$_{0.67}$As layers
(see sample details in \cite{Butov1999} where the same sample was
studied). The electric field across the sample was controlled by an
applied gate voltage $V_g$. The excitons were photoexcited by a cw
633 nm HeNe laser or tunable Ti:Sapphire laser focused to a spot
$\sim 5 \mu$m in diameter. The excitation was circularly polarized
($\sigma^+$). The emission images in $\sigma^+$ and $\sigma^-$
polarizations were taken by a CCD with an interference filter $800
\pm 5$ nm, which covers the spectral range of the indirect excitons.
The spatial resolution was 1.4 micron. The spectra were measured
using a spectrometer with resolution 0.3 meV.

\begin{figure}
\begin{center}
\includegraphics[width=5.3cm]{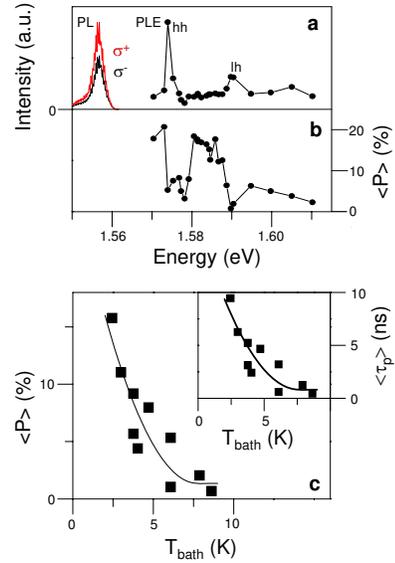}
\caption{\label{fig:fig2}(a) PLE spectrum and polarization resolved
PL spectra of indirect excitons. (b) Spatially and spectrally
averaged PL polarization of indirect excitons $\left<P\right>$;
$V_g=-1.1$ V, $T_{bath}=1.7$ K, $P_{ex}=5\mu$W, PL is measured with
laser excitation energy $E_{ex}=1.573$ eV. (c) $\left<P\right>$ and
(inset) the polarization relaxation time
$\left<\tau_p\right>=\left<\tau_r\right>\left<P\right>/(1-\left<P\right>)$
as a function of $T_{bath}$; $V_g=-1.4$ V, $P_{ex}=10\mu$W, and
$E_{ex}=1.572$ eV.}
\end{center}
\end{figure}

Figure 2a shows the photoluminescence excitation (PLE) spectrum and
polarization-resolved photoluminescence (PL) spectra of the indirect
excitons. PLE reveals two peaks in the exciton absorption which
correspond to the heavy-hole (hh) and light-hole (lh) direct
excitons. The emission of the hh indirect exciton is observed at an energy below
the hh direct exciton by $\sim eF_zd$, where $F_z$ is the electric
field and $d$ is the distance between the electron and hole layers
in CQW (corrections due to the interaction are described in
\cite{Butov1999b}). Figure 2b shows degree of the circular
polarization of PL of indirect excitons. No polarization was
observed at strongly nonresonant excitation by HeNe laser $\sim 400$
meV above the indirect exciton. However, a substantial polarization
of PL of indirect excitons was observed when the excitation was
close to the hh direct exciton (Fig. 2b) at low temperatures (Fig.
2c). The polarization is reduced when the lh exciton (with $m_h=\pm
1/2$) is excited (Fig. 2a,b), consistent with the data in Ref.
\cite{Damen1991} and related to increased spin phase space (e.g. for
the exciton with $s_z=+1/2$ and $m_h=+1/2$ the hole spin relaxation
$m_h=+1/2 \to m_h=-3/2$ changes the emission polarization from
$\sigma^+$ to $\sigma^-$).

The measured degree of the circular polarization $P$ (Fig. 2c) and
the measured exciton lifetime $\tau_r$ \cite{Butov1999} give an
opportunity for estimating the polarization relaxation time $\tau_p$
using the rate equations \cite{Maialle1993,Vinattieri1994}, which
result to $\tau_p=\tau_rP/(1-P)$. The estimate for $\tau_p$ is
presented in the inset to Fig. 2c (the dependence of $\tau_r$ on
parameters, such as temperature in Fig. 2 and excitation power in
Fig. 4, is taken into account in the estimates). The depolarization
time of the emission of indirect excitons reaches 10 ns (Fig. 2c),
orders of magnitude higher than that of the direct excitons in
single QW \cite{Maialle1993,Vinattieri1994}. The depolarization time
of the emission of indirect excitons $\tau_p$ is comparable to their
lifetime $\tau_r$, which is in the range of tens of ns
\cite{Butov1999}.

\begin{figure}
\begin{center}
\includegraphics[width=7cm]{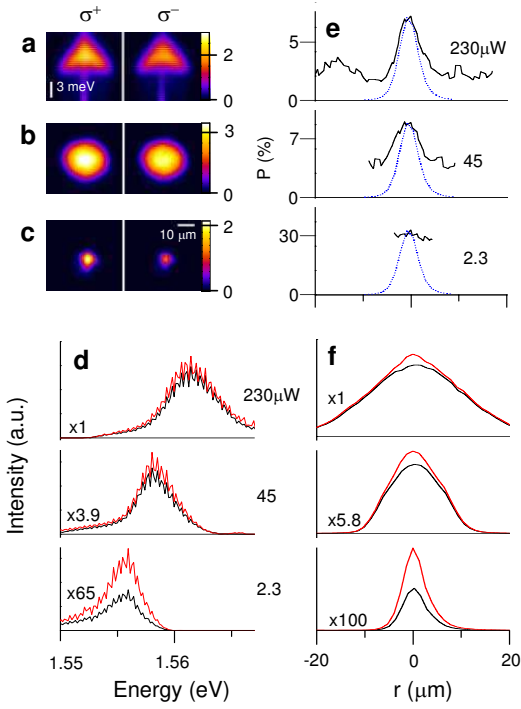}
\caption{\label{fig:fig3}(a) Energy-$x$ images of the PL intensity
of indirect excitons in $\sigma^+$ and $\sigma^-$ polarizations;
$V_g=-1.1$ V, $E_{ex}=1.572$ eV, $P_{ex}=140\mu$W. $x-y$ images of
the PL intensity of indirect excitons in $\sigma^+$ and $\sigma^-$
polarizations for (b) $P_{ex}=15\mu$W and (c) $P_{ex}=4.7\mu$W;
$V_g=-1.1$ V, $E_{ex}=1.582$ eV. (d) PL spectra of indirect excitons
at the center of the exciton cloud ($r=0$) in $\sigma^+$ (red) and
$\sigma^-$ (black) polarizations. The estimated exciton density at
\(r=0\) for \(P_{ex}=\) 2.3\(\mu W\), 45\(\mu W\), and 230\(\mu W\)
is \(9\cdot 10^8\), \(2\cdot 10^{10}\), and \(4\cdot 10^{10}cm^{-2}\),
respectively. The density estimation is described in the text.
$E_{ex}=1.572$ eV. (e) PL polarization as a function of $r$
for the same $P_{ex}$ as in (d). The profile of the bulk emission,
which presents the excitation profile, is shown by dotted lines. (f) PL
intensity of indirect excitons in $\sigma^+$ (red) and $\sigma^-$
(black) polarizations as a function of $r$ for the same $P_{ex}$ as
in (d,e). $T_{bath}=1.7$ K.}
\end{center}
\end{figure}

Characteristic polarization-resolved energy-x and x-y images are shown in
Fig. 3a-c. Figure 3d shows the density dependence of the spectra of the
indirect excitons at the center of the exciton cloud in $\sigma^+$
and $\sigma^-$ polarizations. The polarization and estimate for the
polarization relaxation time is shown in Figs. 4a,b. The exciton
density $n$ was estimated from the energy shift $\delta E$ as $n =
\epsilon \delta E /(4\pi e^{2} d)$ \cite{Ivanov2006} (note that a
higher estimate for $n$ was suggested in Ref. \cite{Schindler2008}).
The polarization degree of the exciton emission (Figs. 3d, 4a) and
the polarization relaxation time (Fig. 4b) reduce with increasing
density. (Note that no increase of $P$ with $n$ such as reported in
Ref. \cite{Larionov2000} was observed in the present experiments.)
Note also that this density dependence is consistent with the observed
reduction of polarization when the excitation energy corresponds to hh
exciton, 1.572 eV, where an increased absorption results in a high
density (Fig. 1b).

\begin{figure}
\begin{center}
\includegraphics[width=5.7cm]{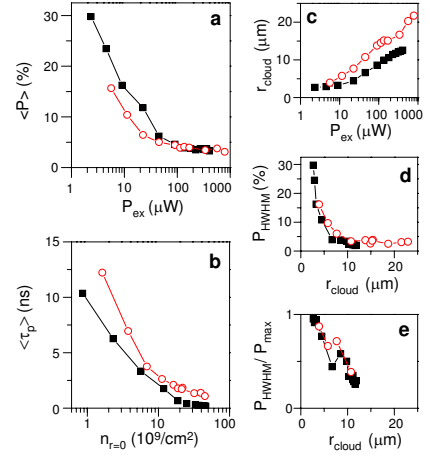}
\caption{\label{fig:fig4} (a) Spatially and spectrally averaged PL
polarization as a function of excitation power. (b) Spatially and
spectrally averaged polarization relaxation time as a function of
density. (c) HWHM of the cloud of indirect excitons $r_{cloud}$ as a
function of excitation power. (d) Polarization at the HWHM of the
exciton cloud $P_{HWHM}$ as a function of $r_{cloud}$. (e)
$P_{HWHM}/P_{max}$ as a function of $r_{cloud}$. $V_g=-1.1$V
(squares), $-1.6$V (circles). $T_{bath}=1.7$ K.}
\end{center}
\end{figure}

The increase of applied gate voltage $|V_g|$ results in the
reduction of the electron-hole overlap and, in turn, increase of
both $\tau_r$ and $\tau_{ex}$. The former was measured in Ref.
\cite{Butov1999} for the sample. The polarization
$P=\tau_p/(\tau_p+\tau_r)$ reduces with reducing electron-hole
overlap (i.e. at higher $|V_g|$), see the data for $V_g=-1.1$ and
$-1.6$V in Fig. 4a. Such dependence indicates a slower increase of
$\tau_p$ than $\tau_r$ with increasing $|V_g|$. This is consistent
with the saturation of $\tau_p$ in the limit of vanishing
electron-hole overlap where $\tau_{ex}$ becomes large compared to
$\tau_e$ and $\tau_h$ so that
$\tau_p^{-1}=2(\tau_e+\tau_h)^{-1}+\tau_{ex}^{-1}$ is determined by
$\tau_e$, for which no substantial dependence on the electron-hole
overlap is expected \cite{Maialle1993,Vinattieri1994}. (This is also
consistent with the fact that essentially no polarization of the
emission of the indirect excitons was observed in another studied
GaAs/AlGaAs CQW sample with 15 nm QWs; in this sample, the larger
separation between the electron and hole layers results to a much
smaller electron-hole overlap, as revealed by a considerably longer
$\tau_r$, in the range of several $\mu s$.) Figure 4b indicates that
$\tau_p=\tau_rP/(1-P)$ still increases with reducing electron-hole
overlap (i.e. at higher $|V_g|$).

The radius of the exciton cloud is essentially equal to the
excitation spot radius at low densities and increases at high
densities, Figs. 3b, c, f, 4c. Such behavior was observed earlier and
attributed to the exciton localization due to in-plane disorder at low
densities and exciton delocalization and transport away from the
excitation spot at high exciton densities when the disorder is
screened by the repulsive interaction between the indirect excitons
\cite{Butov2002,Ivanov2006}. Previous studies indicate that in this
range of temperatures and densities, exciton transport in the CQW
structure can be described by drift and diffusion \cite{Ivanov2006}.

Figures 3f and 3e show the polarization-resolved PL intensity and
polarization $P$, respectively, as a function of distance from the
center of the laser excitation spot. The polarization at the
half-width-half-maximum (HWHM) of the exciton cloud $P_{HWHM}$ (Fig.
4d) and the ratio $P_{HWHM}/P_{max}$ (Fig. 4e) show that the
polarization is observed up to several microns away from the origin
which gives the length scale for the exciton spin transport in the
structure. This length scale is large enough to allow studying the
basic properties of the exciton spin transport by the optical
experiments (it is beyond the diffraction-limited spatial
resolution, which is 1.4 microns in the present work). It is also
large enough to allow studying spin-polarized exciton gases in
microscopic patterned devices, e.g. in in-plane lattices
\cite{Hammack2006}, which period can be below a micron. Furthermore,
the length scale for exciton spin transport is comparable to the
scale of the excitonic devices such as the excitonic transistors and
circuits \cite{High2007,High2008} (the distance between source and
drain for an excitonic transistor was 3 $\mu m$ in Refs.
\cite{High2007,High2008}; however, it is expected that the
dimensions can be reduced below 1 $\mu m$ by using e-beam
lithography). This may allow creating spin-optoelectronic devices
where the spin fluxes of excitons will be controlled in analogy to
the control of the fluxes of unpolarized excitons in Refs.
\cite{High2007,High2008}.

A quick reduction of the polarization is observed in the vicinity of
the excitation spot, followed by the tail with a weak spatial
dependence, Fig. 3e,f. Since the lifetime $\tau_r$ and polarization
relaxation time $\tau_p$ of the indirect excitons are comparable,
the quick reduction of the polarization on a length scale smaller
than the diffusion length $\left(\sqrt{D_s\tau_p} <
\sqrt{D\tau_r}\right)$ indicates that the exciton spin diffusion
coefficient $D_s$ near $r=0$ is smaller than the exciton diffusion
coefficient $D$. Note parenthetically that in the case of electron
transport, the spin Coulomb drag results to a smaller value of $D_s$
compared to $D$ \cite{Amico2001,Weber2005}. Note also that the
exciton temperature is higher in the vicinity of the excitation spot
\cite{Ivanov2006} and this contributes to a quicker reduction of the
polarization there according to the temperature dependence in Fig. 2c.
Theory of spin relaxation and transport of indirect excitons, which
can be compared to the experimental data, has yet to be developed.

In conclusion, the spatially resolved circular polarization of the
emission of long-life indirect excitons in CQW structures was
studied. The polarization relaxation time of the emission of the
indirect excitons exceeds that of regular direct excitons by orders
of magnitude and reaches ten ns. The polarization degree and the
polarization relaxation time reduce with increasing density and
temperature. Spin transport of the indirect excitons was observed.
The exciton spin transport originates from the long spin relaxation
time and long lifetime of the indirect excitons.

This work is supported by DOE. We thank Misha Fogler, Lu Sham, and
Congjun Wu for discussions.

\end{document}